\documentclass[iop]{emulateapj}
\usepackage{epsfig}

\usepackage[usenames,dvips]{color}

\shorttitle{NuSTAR Observations and modeling of PSR J1023+0038}
\shortauthors{Li et al.}

\newcommand{\flux}{\,erg\,cm$^{-2}$\,s$^{-1}$}

\newcommand{\cm}{\,cm$^{-2}$}
\newcommand{\nh}{$N_\mathrm{H}$}

\newcommand{\chandra}{{\it Chandra}}

\newcommand{\xmm}{{\it XMM-Newton}}
\newcommand{\swift}{{\it Swift}}

\newcommand{\nustar}{{\it NuSTAR}}
\newcommand{\psr}{J1023}

\begin{document}
\slugcomment{Accepted for publication in ApJ}
\title{NuSTAR observations and broadband spectral energy distribution modeling of the millisecond pulsar binary PSR~J1023+0038}

\author{K.~L. Li\altaffilmark{1}, A.~K.~H.~Kong\altaffilmark{1}, J. Takata\altaffilmark{2}, K.~S. Cheng\altaffilmark{2}, P.~H.~T. Tam\altaffilmark{1}, C.~Y. Hui\altaffilmark{3}, Ruolan Jin\altaffilmark{1}
}

\affil{$^1$ Institute of Astronomy and Department of Physics, National Tsing Hua University, Hsinchu 30013, Taiwan; lilirayhk@gmail.com, akong@phys.nthu.edu.tw}
\affil{$^2$ Department of Physics, University of Hong Kong,
Pokfulam Road, Hong Kong; takata@hku.hk} 
\affil{$^3$ 
Department of Astronomy and Space Science, Chungnam National University, Daejeon, Republic of Korea}  



\begin{abstract}
We report the first hard X-ray (3--79 keV) observations of the millisecond pulsar (MSP) binary PSR J1023+0038 using {\it NuSTAR}. This system has been shown transiting between a low-mass X-ray binary (LMXB) state and a rotation-powered MSP state. The {\it NuSTAR} observations were taken in both LMXB state and rotation-powered state. The source is clearly seen in both states up to $\sim 79$ keV. During the LMXB state, the 3--79 keV flux is about a factor of 10 higher that in the rotation-powered state. The hard X-rays show clear orbital modulation during the X-ray faint rotation-powered state but the X-ray orbital period is not detected in the X-ray bright LMXB state. In addition, the X-ray spectrum changes from a flat power-law spectrum during the rotation-powered state to a steeper power-law spectrum in the LMXB state. We suggest that the hard X-rays are due to the intra-binary shock from the interaction between the pulsar wind and the injected material from the low-mass companion star. During the rotation-powered MSP state, the X-ray orbital modulation is due to Doppler boosting of the shocked pulsar wind. At the LMXB state, the evaporating matter of the accretion disk due to the gamma-ray irradiation  from the pulsar stops almost all the pulsar wind, resulting the disappearance of the X-ray orbital modulation.
\end{abstract}


\section{Introduction}
Millisecond pulsars (MSPs) are widely believed to be the descent of low-mass X-ray binaries (LMXBs). Based on current theoretical models, the rotating neutron star in a LMXB can be spun up to millisecond periods via accretion from the low-mass companion which transfers mass and angular momentum to the pulsar (e.g., Alpar et al. 1982). The first evidence of this ``recycling'' scenario comes from the discovery of accreting millisecond X-ray pulsars in LMXBs (e.g., Wijnands \& van der Klis 1998). When the accretion stops and the accretion disk is removed, the system will become a MSP powered by rotation. It is still a puzzle how and when this process happens. Possible models include pulsar wind ablation (Wang et al. 2009), propeller effect (Romanova et al. 2009), and $\gamma$-ray heating from the MSP (Takata et al. 2010).

To investigate the evolutionary process of MSPs and LMXBs, one has to look for a system that shows both rotation-powered MSP and LMXB behaviors. The discovery of the MSP/X-ray binary PSR J1023+0038 (hereafter J1023) is key to understanding the state transition between a LMXB and a MSP. J1023 was first suggested as a magnetic cataclysmic variable (Bond et al. 2002) and was subsequently identified as a LMXB candidate (Thorstensen \& Armstrong 2005; Homer et al. 2006). Radio observations then show that J1023 is a 1.69-ms pulsar in a 4.8-hr binary orbit with a $\sim 0.2 M_\odot$ companion star (Archibald et al. 2009), establishing a link between a MSP and a LMXB. Interestingly, the system shows clearly accretion disk signature before 2002 (Wang et al. 2009), but the disk has disappeared since then. This is a direct evidence for a state transition from a LMXB to a MSP. More recently, the X-ray emission of J1023 has increased by a factor of about 20 (Kong 2013; Patruno et al. 2014; Takata et al. 2014) since 2013 October and the radio pulsation has disappeared after 2013 mid-June (Stappers et al. 2014). Meanwhile, the UV emission has brightened by 4 magnitudes (Patruno et
al. 2014) and the accretion disk has reemerged based on optical spectroscopy (Takata et al. 2014), indicating that J1023 has switched from a MSP to a LMXB. It is now clear that J1023 underwent two state transitions between a LMXB and a MSP. It is worth noting that there are two similar systems, IGR J18245--2452 (Papitto et al. 2013) and XSS J12270--4859  (Bassa et al. 2014), that also show similar state transitions. 
Such state change events have proven solidly that MSPs are evolved from LMXBs.

J1023 has been observed in X-ray with \chandra, \xmm, and \swift\ in the past decade. In particular, the \xmm\ observations reveal the X-ray pulsation as well as the X-ray orbital modulation (Archibald et al. 2010; Tam et al. 2010) when the source is in the X-ray faint rotation-powered state. During the recent X-ray bright LMXB state, the X-Ray Telescope (XRT) onboard \swift\ detected the soft (0.3--10 keV) X-ray brightening (Kong 2013; Patruno et al. 2014; Takata et al. 2014) while the Burst Alert Telescope (BAT) failed to detect the hard (15--50 keV) X-rays (Stappers et al. 2014). J1023 also has detectable $\gamma$-rays (200 MeV--20 GeV) with the {\it Fermi Gamma-ray Space Telescope} in both rotation-powered and LMXB states (Tam et al. 2010; Takata et al. 2014).

\begin{figure*}[t]
\centering
\includegraphics[width=170mm]{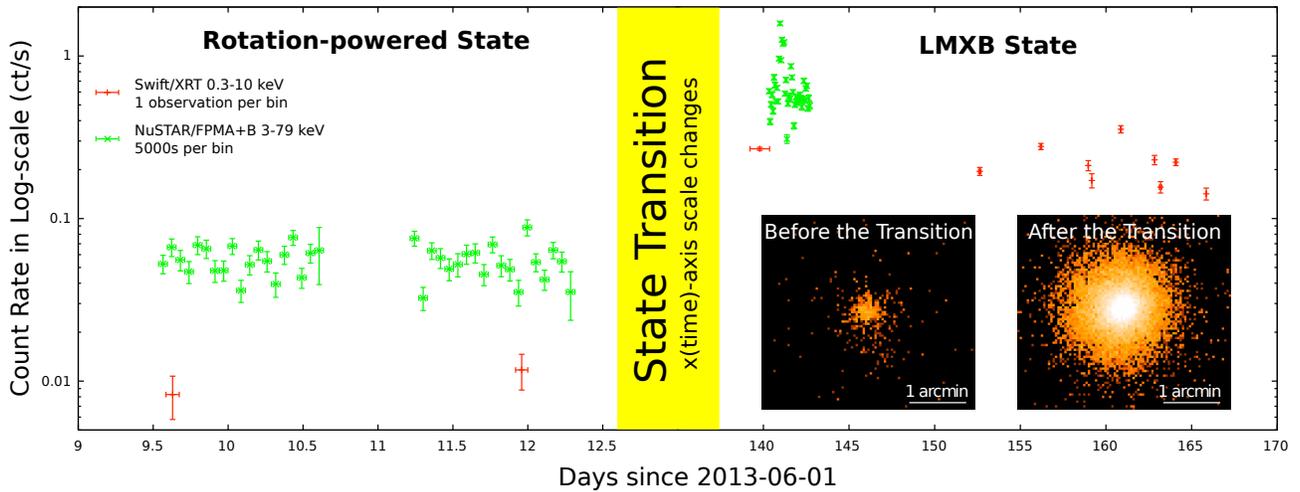}
\caption{Long-term X-ray lightcurves of \textit{NuSTAR}/FPMA/B (green) and \textit{Swift}/XRT (red; see Takata 2014 for details) before/after the state transition with different time-axis scale. The inset box shows the \textit{NuSTAR}/FPMA+B images before/after the transition. The intensity ratio between the two images is about 10. 
}
\label{fig:lc}
\end{figure*}

In this paper, we report the first \nustar\ hard X-ray (3--79 keV) observations of J1023 taken before and after its recent MSP/LMXB state change, and discuss the origin of the X-ray emission in the context of an intrabinary shock model.

\section{Observations and Data Reduction}

\textit{NuSTAR} is the first focusing hard X-ray telescope that is sensitive to 3--79 keV (Harrison et al. 2013). It observed \psr\ four times with the on board focal plane modules A and B (FPMA/B) in mid-June and mid-October 2013, during which the MSP was in the rotation-powered (X-ray low state) and LMXB (X-ray high state) states, respectively (Patruno et al. 2014; Takata et al. 2014). Figure 1 shows the long-term lightcurves of J1023 observed with \nustar\ and \swift\ since 2013 June 1.
The total exposure time for the first three observations taken in 2013 June is 94.1 ks, which is roughly the same as the last observation in 2013 October with an integration time of 94.3 ks. 
We downloaded data products of the observations including the housekeeping, auxiliary, and cleaned/calibrated events (i.e., Level 2) files from the \texttt{HEASARC} data archive. 
The X-ray counterpart of \psr\ is clearly detected within the \textit{NuSTAR}'s absolute astrometric accuracy (5\arcsec at 90\% confidence; Alexander et al. 2013) in all observations and the source is significantly brighter after the state transition (see the inset box of Figure \ref{fig:lc}). 

Scientific data extractions are performed using the \texttt{HEASARC}'s \texttt{HEASoft} multi-mission software (version 6.15.1) together with NuSTARDAS v1.3.1 and the updated \nustar\ calibration database (CALDB version 20131223) by following the instructions outlined in the \textit{NuSTAR} Data Analysis Software Guide\footnote{http://heasarc.gsfc.nasa.gov/docs/nustar/analysis/nustardas\_swguide\_v1.5.pdf}. The X-ray spectra, the corresponding response matrices, and lightcurves used in this work are produced by a \textit{NuSTAR}-specific task \texttt{nuproducts} by keeping most of the parameters of the task default, but using different PI-channel ranges of interest (will be discussed $\S3.2$) and turning the barycentric correction option on. For the extraction regions, we choose different sizes of source regions based on the source count rate of the epoch. For the low-state observations, a circular source region of $30\arcsec$ radius and a source-free annulus background region of $80\arcsec$/$160\arcsec$ inner/outer radii around the source are applied. For the high-state observation, we used a larger circular source region of $50\arcsec$ radius to encompass more useful source photons. As the source is close to the contact between the chips (each FPM consists of a $2\times2$ Cadmium-Zinc-Tellurium detector array), we select three circular regions with radii of $50\arcsec$ (identical to the source region) at similar chip positions as the background to minimize the inconsistency between the source/background samplings. Unless otherwise mentioned, the uncertainties listed in this paper are in 90\% confidence level.

\section{Data Analysis and Results}
\subsection{Spectral Analysis}

\begin{figure*}
\centering
\includegraphics[width=170mm]{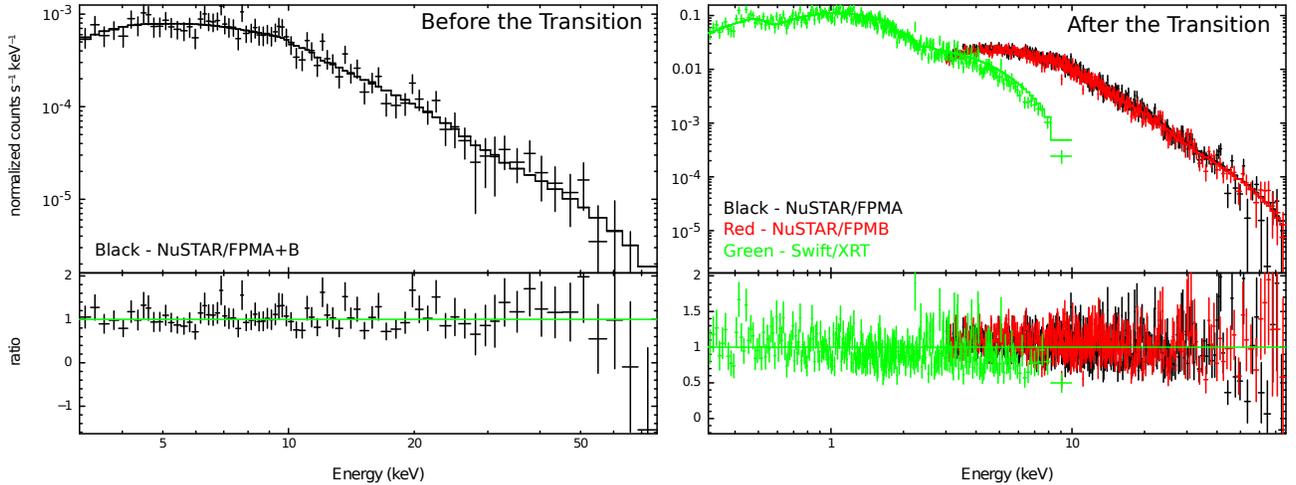}
\caption{\textit{NuSTAR}/FPMA+B (and \textit{Swift}/XRT) spectra of \psr\ during the rotation-powered state (left) and the LMXB state (right). Both datasets can be fit with a simple power-law. The X-ray flux during the LMXB state is about 10 higher than the quiescent value and the photon index is significantly softer.
}
\label{fig:spec}
\end{figure*}

We performed X-ray spectral model fittings for all four datasets with \texttt{XSPEC}. Since the numbers of detected photons are small for the first three observations taken in the X-ray low state, we therefore stacked the three datasets for spectral analysis. As \textit{NuSTAR} observed \psr\ roughly simultaneously with \textit{Swift}/XRT (Figure \ref{fig:lc}), we also included the high-state XRT spectrum to have a much wider energy range (0.3--79 keV) for the spectral fittings while the low-state XRT spectrum is not used because of its very limited photon statistics.
Details of the \textit{Swift}/XRT spectrum/lightcurve used in this analysis have been discussed in our previous paper (Takata et al. 2014). All spectra were grouped with at least 25 counts per spectral bin to allow $\chi^2$ statistics.

During the quiescence, the spectrum can be fit with an absorbed power-law model ($\chi^2/dof=56.24/76$; Figure \ref{fig:spec}) with a column density \nh\ $<3.4\times10^{22}$\cm, a photon index of $\Gamma=1.20^{+0.14}_{-0.09}$, and an observed 3--79 keV flux of $F_\mathrm{obs}=2.89^{+0.26}_{-0.5}\times10^{-12}$\flux\ (Figure \ref{fig:spec}), which are consistent to the spectral fitting result of the 83.1~ks \textit{Chandra} observation taken on 2010 March 24 (Bogdanov et al. 2011). 
The absorption parameter is poorly constrained since the photoelectric absorption is only sensitive to soft X-rays (i.e., less than 2~keV), instead of the hard \textit{NuSTAR}'s energy band. Nonetheless, for the same reason, the ambiguous column density does not significantly affect our result. In fact, the observed and the absorption corrected fluxes of the fits are almost the same. 
For the \textit{NuSTAR} and \textit{Swift} observations taken during the LMXB state, an absorbed power-law is still the best model yielding a chi-square statistic value of $\chi^2/dof=1296.29/985$ with the best-fit values of \nh\ $=4.9^{+0.8}_{-0.7}\times10^{20}$\cm, $\Gamma=1.63\pm0.01$, and $F_\mathrm{obs}=2.67^{+0.05}_{-0.04}\times10^{-11}$\flux\ (Figure \ref{fig:spec}). 
All parameters including the column density are greatly constrained owing to a much higher signal-to-noise ratio and a wider energy range (i.e., 0.3--79~keV) of the resultant spectrum. 
Comparing with the quiescent observations, the photon index is significantly softer and the observed flux is about 10 times larger after the state transition, which are consistent with the previous \textit{Swift}/XRT observations (Patruno et al. 2014; Takata et al. 2014). 

Besides a simple power-law model, we tried to add a high-energy exponential cut-off component to improve the spectral fitting results of the \textit{NuSTAR} data. However, based on the $\chi^2$ values, the improvement is insignificant for the high-state spectrum (the F-test probability is 59\%) and there is no improvement for the low state. Furthermore, the best-fit cut-off energy is about 800~keV for the high-state and is pegged to the maximum allowed value (i.e., 5000~keV) for the low state. In addition to the unreasonable large uncertainties of the cut-off energies (i.e. a few times of the best-fit value), both cut-off energies are significantly larger than the energy range of NuSTAR strongly suggesting that the fits are unreliable. Therefore, we conclude that a simple power-law model is the best spectral model for the data.

\subsection{Temporal Analysis}

\begin{figure*}
\centering
\includegraphics[width=165mm]{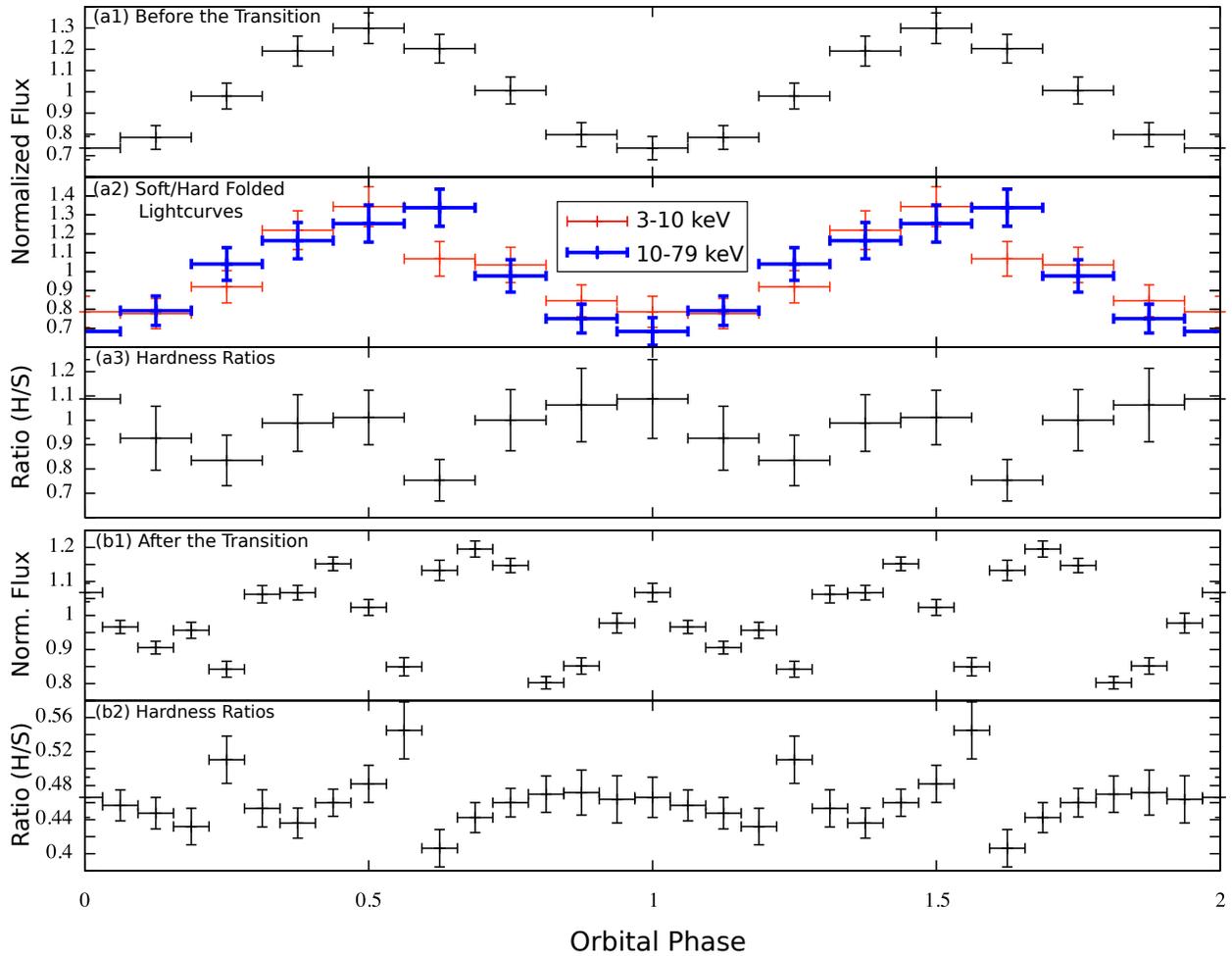}
\caption{The figure shows the barycentered \nustar\ lightcurves folded at the binary period ($P_\mathrm{orb}=17115.512$ s) with the phase zero defined as the inferior conjunction. Top panel: full band (3--79~keV) lightcurves during the rotation-powered state. Second panel: hard (blue/thick; 10--79~keV) and soft band (red/thin; 3--10~keV) band lightcurves. Third panel: ratios between hard and soft band data points. Fourth panel: full band (3--79~keV) lightcurves during the LMXB state. Fifth panel: ratios between hard and soft band data points.
}
\vspace*{9mm}
\label{fig:orbit}
\end{figure*}

To search for the hard X-ray orbital modulation of \psr\, we use the \texttt{XRONOS} package to analyse the \textit{NuSTAR} low/high-state lightcurves that cover about 5.5 orbital cycles. The barycentric correction is applied to the lightcurves as mentioned in $\S2$. We also set the bin size of the lightcurves to 10 s, which is fine enough to search for the 4.8-hr orbital period (Thorstensen \& Armstrong 2005; Archibald et al. 2009). 
Since the source lightcurves generated by \texttt{nuproducts} are not background subtracted, we do the background subtraction by \texttt{lcmath} with the background lightcurves scaled by the ratio of the integrated effective exposures within the source and background regions. The contributions of the background are just 2\% and 0.03\% of the total count rate during the low and high states, respectively. For the timing ephemeris, high-precision measurements of the binary parameters measured by the GBT, the Arecibo Observatory, and Parkes Observatory (Archibald et al. 2009) are adopted, however, the orbital period derivatives are not considered for simplicity. The top (a1) and fourth (b1) panels of Figure \ref{fig:orbit} show the folded low/high-state lightcurves by \texttt{efold} with the orbital phase zero defined as the inferior conjunction (i.e., the companion is between the pulsar and the observer). 
For the quiescent folded lightcurve, it can be well described by a sinusoidal function with a peak amplitude of about 30\% of the average intensity (significance is $\sim10\sigma$).
With the minimum intensity occurring at an orbital phase $\phi$ of 0 and 1, and the maximum at $\phi=0.5$, it is unambiguously in phase with the optical modulation (Thorstensen \& Armstrong 2005) and the soft X-ray (0.3--8~keV) modulation measured by \textit{Chandra} (Bogdanov et al.
2011). We split the \nustar\ data into soft (3--10~keV) and hard (10--79~keV) energy band to search for an orbital-phase dependent spectral variability that is absent in the \textit{Chandra} data (Bogdanov et al. 2011). No statistically significant spectral variation has been found from the hard/soft ratio against orbital phase (Figure \ref{fig:orbit}) that confirms previous \textit{Chandra} result. 

\begin{figure*}
\centering
\includegraphics[width=165mm]{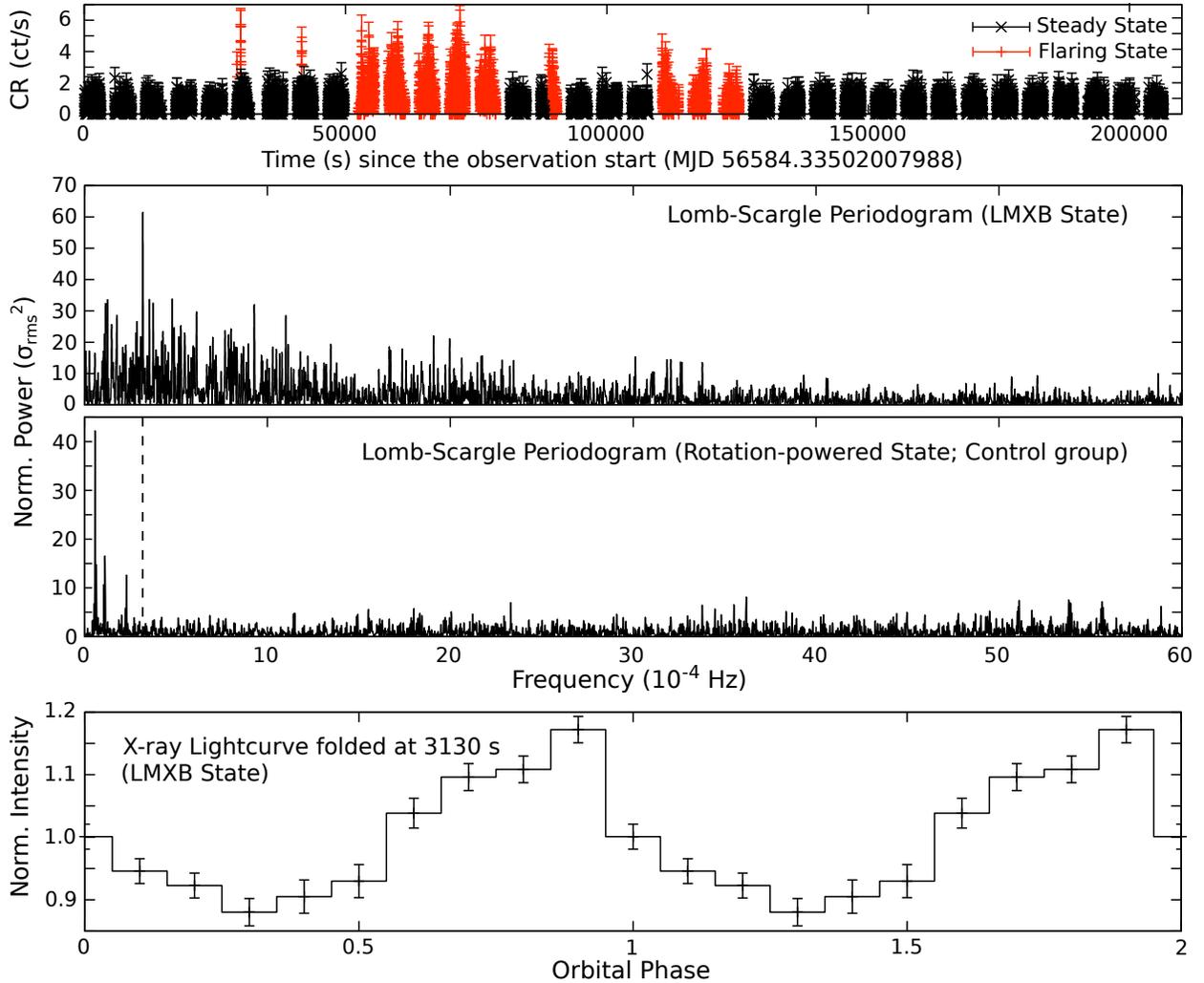}
\caption{The first panel shows the \textit{NuSTAR} high-state lightcurve of \psr\ with a binning factor of 10 s. Data in red indicate the flaring states while the black ones indicate the steady states that are used for the Lomb-Scargle analysis. 
The Lomb-Scargle periodogram of the high-state data is shown in the second panel with a control group of low-state data in the third panel of which the three highest peaks correspond to the orbital modulation and its harmonic frequencies. A peak of normalized power $\sim\,60\sigma_\mathrm{rms}^2$ (false-alarm probability of the null hypothesis $\approx 10^{-23}$) at a frequency of $f=(3.195\pm0.002)\times10^{-4}$ Hz (i.e., $P=3130\pm2$ s) is clearly seen in the high-state but is invisible during the low-state (indicated by a dashed line), suggesting that it is not systematic. The last plot is the high-state lightcurve folded at 3130 s (with an arbitrary phase zero) indicating a quasi-sinusoidal brightness variation. 
}
\label{fig:powspec}
\end{figure*}

For the high-state folded lightcurve, although a high variability is recorded with an extreme $\chi^2$ value of $\sim600$ ($dof=15$), the curve is not compatible with a sinusoidal function and seemingly nonperiodic indicating that the variability
at this frequency is likely caused by random fluctuations instead of a combination of periodic signals. 
Interestingly, we tried to fold it with an arbitrary period but the high variability is still present, strongly suggesting that the variability is likely to be dominated by other factors instead of the viewing-angle at different phases. We note that soft X-ray rapid flickering is also seen by \textit{Swift} during the X-ray high state (Patruno et al. 2014). We also performed a Lomb-Scargle analysis on the high-state data to search for other periodicity. After removing visually detected flaring states (timescales of the flares $\approx100-1000$ s) as well as some individual flares for which the count rate is higher than 3 counts per second in a 10-s bin (see the first panel of figure $\ref{fig:powspec}$ for details), a strong peak with a normalized power of $\sim60\,\sigma_\mathrm{rms}^2$ is clearly seen in the periodogram at $f=(3.195\pm0.002)\times10^{-4}$ Hz (equivalent to a period of $P=3130\pm2$ s) in which the uncertainty is estimated by a bootstrapping technique with 5000 trials. The periodicity has not been detected by \textit{Swift}/XRT during the earlier target-of-opportunity observations (Takata et al. 2014) probably due to the limited data quality of \textit{Swift} as the XRT data are fragmented into observations with several different short exposures. 
A similar analysis is also performed on the \textit{NuSTAR} low-state data as a control to test whether the periodicity is a intrinsic property of the pulsar binary or an instrumental pattern. Except three peaks corresponding to the orbital modulation of \nustar, and its harmonic frequencies, the low-state periodogram is basically featureless suggesting that the 3130~s periodicity is not systematic. Moreover, a lightcurve folded at the 3130-s period produces a quasi-sinusoidal modulation that is unlikely to be produced randomly (Figure $\ref{fig:powspec}$). Indeed, the same modulation can also be produced by the soft and hard bands data without any significant spectral variation. 
Besides to the 3130-s periodicity, the power of the periodogram is significantly higher than the control one (i.e., up to $30\,\sigma_\mathrm{rms}^2$) around the frequencies from $10^{-4}$ to $3\times10^{-3}$~Hz implying variabilities with timescales ranging from a few minutes to a few hours are strong during the LMXB state. Short-term peculiar dips with variable lengths of 200--800 s are detected, which confirms the previous \textit{Swift}/XRT finding of the low-flux intervals ranging from 200 to 550 s (Takata et al. 2014).  

\section{Discussion}

We have found hard X-ray (3--79 keV) emission from the MSP/LMXB J1023 using \nustar. The hard X-rays were seen in both rotation-powered and LMXB states. At the distance of 1.3 kpc, the 3--79 keV luminosity is $5.8\times10^{32}$ erg s$^{-1}$ and $5.4\times10^{33}$ erg s$^{-1}$ during the rotation-powered state and the LMXB state, respectively. It is therefore not surprising that \swift/BAT could not detect J1023 (Stappers et al. 2014) because the sensitivity limit is at least two orders of magnitude higher. Comparing to previous soft X-ray observations, the \nustar\ spectra are essentially the non-thermal extension from the soft X-rays. In both states, the broad band spectrum up to 79 keV can be fit with a simple power-law with $\Gamma=1.20$ ($\Gamma=1.63$) in the low (high) state, which is more or less the same as the spectral shape derived from soft X-ray observations (Archibald et al. 2010; Bogdanov et al. 2011; Patruno et al. 2014; Takata et al. 2014). It is worth noting that a broken power-law model with a cutoff energy of 1.8 keV (Tam et al. 2010), or a thermal emission component from the neutron star (Bogdanov et al. 2011) can also describe the spectral shape during the rotation-powered state. However, \nustar\ is not sensitive to these soft X-ray features.

\begin{figure*}
\begin{center}
\includegraphics[width=175mm]{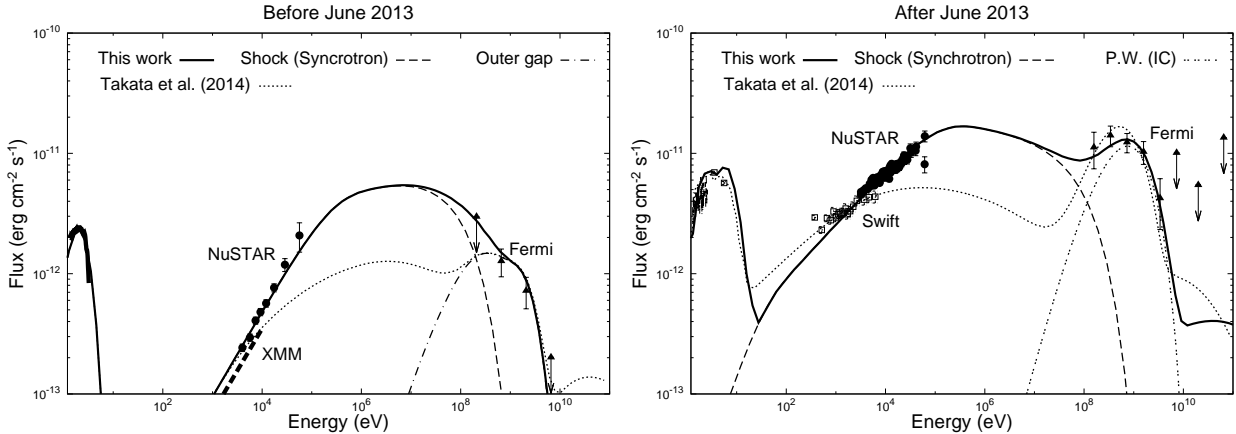}
\caption{Multi-wavelength spectra before (left) and after (right) June 2013. 
The solid lines and dotted lines represent the calculated spectra 
in this work and in Takata et al. (2014), respectively. The dashed-dotted line 
in the left panel shows the spectrum of the outer gap emissions, and 
the double-dotted line in right panel shows the inverse-Compton 
spectrum of the cold-relativistic pulsar wind.  
For shock emissions (dashed-lines), we assume the shock is 
located at  $r_s=3\times 10^{10}$cm and assume that 
10\% and 70\% of the pulsar wind is stopped by the shock before and after 
June 2013, respectively.  In addition, the power law indexes of the shocked 
particles are assumed to be $p=1.5$ and $p=2$ before and after June 2013, 
respectively. A more detail calculation method 
can be seen in Takata et al. (2014). }
\label{fit}
\end{center}
\end{figure*} 

In terms of timing behavior, we found that the hard X-rays show the 4.8-hr orbital modulation during the low state while it disappears in the high state (Fig. 3). We will explain the X-ray orbital modulation using our model in this section later. Moreover, short-term variabilities on timescales of 100--1000 s are seen in the high state which are consistent with previous soft X-ray observations (Patruno et al. 2014; Takata et al. 2014). We speculate with our intrabinary shock model (Takata et al. 2014) that the observed variability is caused by either perturbation of shock front due to clumpy stellar wind or wind speed variation, or sound propagation in the shock front.
Similar short-term variability has also been found in other two known transitional MSP/LMXB systems, XSS~J12270-4859 (de Martino et al. 2013) and IGR\,J18245Ð-2452 (Ferrigno et al. 2013; Linares et al. 2014), during their X-ray high states. Such a short timescale X-ray variability could be a common signature of MSP/LMXB systems in the LMXB state.
Interestingly, a periodic modulation of 3130 s is found (see Fig. 4) although its nature is not clear.  Additional X-ray observations during the LMXB state are required to better determine the origin.

In the soft X-ray band, the high-energy emission in general is explained by thermal emission from the neutron star surface and non-thermal emission produced in pulsar magnetospheres. Furthermore, X-ray emission in form of X-ray nebulae can be powered by MSPs while intrabinary shock within the binary system can also generate X-rays when the pulsar wind interacts with the materials from the companion star (e.g., Kong et al. 2012; Takata et al. 2014; Hui et al. 2014). It is also suggested that accretion onto the neutron star magnetosphere can produce the non-thermal X-rays (Campana et al. 1998). More recently, it has been proposed that the state transition of another transitional MSP/LMXB IGR\,J18245Ð-2452 is due to transition between magnetospheric accretion and intra-binary shock emission (Linares et al. 2013), or a propeller effect (Papitto et al. 2013). Propeller effect was also suggested to explain the state transition of J1023 (Archibald et al. 2009; Patruno et al. 2014).

In Takata et al. (2014), we propose a theoretical model to
explain the multi-wavelength (from UV to GeV gamma-ray) observations of J1023. 
For the low state (before June 2013), 
we assumed that the momentum ratio between the stellar 
wind and the pulsar wind, $\eta=\dot{M}v_wc/L_{sd}$,
 is much smaller than unity, that is, 
 $\eta=0.3(\dot{M}/5\cdot 10^{15}{\rm g\, s^{-1}})
(v_w/10^8{\rm cm\,s^{-1}})(L_{sd}/5\cdot 10^{34}{\rm erg\, s^{-1}})^{-1}$, 
and therefore  the shock in low state phase (i.e. radio pulsar phase) is 
close to the L1-Lagrangian point. The size of the emission region 
was approximately described by the size of the Roche-lobe.
In such a case, the observed X-ray modulation in the low state will be 
  caused by eclipse of the emission region (see Bogdanov et al. 2011). 
However, we found in Figure~\ref{fit} that 
the predicted flux above 10keV in Takata et al. (2014) 
is about one order of magnitude smaller  than that measured by the \nustar. 
Furthermore, Archibald et al. (2013) argued that it is not clear 
how such a shock geometry, which surrounds the companion, can explain 
the length ($\sim 0.6$ orbital phase centered at inferior conjunction)  
of the observed 350-MHz eclipse, and therefore the shock should cover 
the pulsar. This would indicate that the system is dominated by the outflow from the companion star 
and the mass loss rate is larger than $\sim 2\times 10^{16}{\rm g\, s^{-1}}$. 
 Archibald et al. (2013) argued that the large mass loss rate would be 
 incompatible with the dispersion measure,  and therefore they suggested that 
the  magnetic pressure of the companion star is against the pulsar wind. 
For example, we may assume the stellar magnetic field as 
$B(R)=B_*(R_*/R)^2$ (see  Archibald et al. 2013 and reference therein), 
where $R$ is the distance from the center of the companion, 
and $B_*$ is the magnetic field at the surface.
 In addition,  the radius of the companion star ($R_*$) is estimated as 
the Roche-lobe radius, 
 $R_*\sim 0.462 [q/(1+q)]^{1/3}a\sim 3\times 10^{10}$cm with 
$q=0.2M_{\odot}/1.4M{\odot}\sim 0.14$ being mass ratio 
and $a\sim 1.3\times 10^{11}$cm separation between the two stars. 
The distance to the apex of the shock from the pulsar will be estimated from
 $r_s/a\sim 1/(1+\eta_b^{1/2})$, where $\eta_b=B^2_*R_*^2c/L_{sd}\sim 
5(B_*/10^2{\rm G})^2(R_*/2\cdot 10^{10}{\rm cm})^2(L_{sd}/5\cdot 10^{34} 
{\rm erg s^{-1}})^{-1}$. Hence, the shock could 
 wrap the pulsar  if the magnetic field at the 
stellar surface is larger than $B_*\sim 10^2$G. In the present calculation, 
we use $\eta_b\sim 7$, which corresponds to the shock radius 
$r_s\sim 3\times10^{10}$cm, to fit the X-ray data with the emission model.

  Figure~\ref{fit}  shows
 the results of our revised model fitting for
 the multi-wavelength observations including the \nustar\ data. 
In the present calculation, the fraction of the pulsar wind stopped at the 
shock apex $r\sim r_s$ and the magnetization parameters are used as the 
free parameters.  To fit the X-ray observations of the low state (left panel), we assumed that about 10\% of the pulsar wind is stopped at 
around the shock apex $r\sim r_s$, and we also assumed 
 a magnetization parameter of $\sigma\sim10^{-2}$ at the shock.   
To explain the hard spectrum of the X-ray emissions, 
we assume $p=1.5$ as the power index of the accelerated particles
 at the shock. The predicted luminosity of the emissions 
from the pulsar (outer gap model) and pulsar wind (shock) 
is $L_{gap}\sim  5\times 10^{33}{\rm erg s^{-1}}$ and 
$L_{pw}\sim 4\times 10^{33}{\rm erg s^{-1}}$, respectively.  
We note that the {\it observed} GeV luminosity 
($\sim 5\times 10^{32}{\rm erg s^{-1}}$) 
in low sate is smaller than the predicted luminosity of the outer gap model. 
We expect that our line of sight 
cuts through the edge of the gamma-ray beam and the apparent luminosity 
is smaller than the intrinsic luminosity. 

\begin{figure}
\begin{center}
\includegraphics[height=6cm]{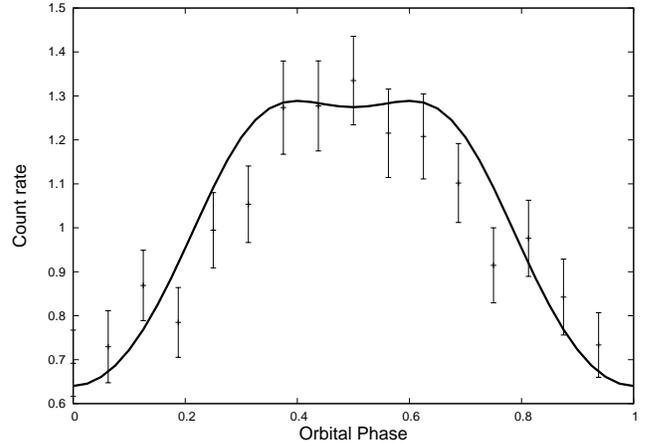}
\caption{The orbital modulation of the X-ray emissions before 
June 2013. The phase zero corresponds to the inferior conjunction, where 
the companion is located between the pulsar and Earth. In the present 
model, the Doppler boosting causes the orbital modulation.
Solid line shows the model light curve with shock geometry characterized by 
the momentum ratio $\eta=7$ and the velocity $v\sim 0.4c$ 
of the shocked pulsar wind.  }
\label{light}
\end{center}
\end{figure}

 In the present model, the shock is located in pulsar side
 and the observed orbital modulation of the X-ray emissions is explained by 
 Doppler boosting due to the finite velocity of the shocked pulsar wind.  
 The Doppler boosting introduces an orbital modulation of 
the emissions that are isotropic in the co-moving frame with the flow. 
Figure~\ref{light} compares the model light curve and the $NuSTAR$ 
observation. For the model, we assume $v=0.4c$ for the velocity
 of the shocked pulsar wind to explain the observed amplitude.

The UV/optical emissions of the high state (after June 2013)
 come from the accretion disk and are also responsible to generate 
additional GeV gamma-rays via inverse
 Compton scattering of the cold-relativistic pulsar wind. 
In this model, we expect that 
the accretion disk in the high state does not come into 
the pulsar magnetosphere, and 
the rotation powered activities of the pulsar are still 
turned on.  For a standard disk model (Frank et al. 2002), the gas pressure  
of  the disk at the distance $r$ from the pulsar is 
\begin{equation}
P_{disk}(r)\sim 3.6\times 10^4\mu^{-1}\alpha^{-9/10}\dot{M}_{16}^{17/20}M_1
r_{10}^{-21/8} {\rm dyn~cm^{-2}},
\end{equation}
where $\mu$ is the average molecular weight, 
 $\alpha$ is the viscosity parameter, $\dot{M}_{16}$ is the mass loss rate 
of the companion in units of $10^{16}{\rm g s^{-1}}$, $M_1$ the neutron star 
mass in units of the solar mass and $r_{10}=(r/10^{10}{\rm cm})$.
 One finds that this gas pressure dominates the pulsar 
wind pressure,  $P_{PW}\sim L_{sd}/4\pi r^2c\sim 1.3\cdot 10^3(L_{sd}/5\cdot 10^{34}{\rm erg s^{-1}})r_{10}^{-2}{\rm dyn~cm^{-2}}$, below  L1-Lagrangian point
 ($r_{L1}\sim 8\times 10^{10}$cm from the pulsar), 
and the pulsar wind cannot stop the migration of the accretion disk.  
Takata et al (2010, 2012) however argued that 
the gamma-rays from the pulsar magnetosphere irradiating the accretion 
disk are absorbed by the disk matter through a pair-creation process 
in the presence of a nucleus (Liang 1999), 
and the energy transfer from the gamma-rays to the disk matters causes 
an evaporation of the disk. Applying a standard 
disk model (Frank et al. 2002), we found that the disk column density below 
$r\sim 3\times 10^9$ cm from the pulsar  is 
 high enough to absorb the gamma-rays. 
In this model, the evaporation rate from the disk is calculated from 
$\zeta L_{sd}=\dot{M}_{ev}v_{es}^2/2$ with $\zeta$ being irradiation efficiency 
and $v_{es}$ escape velocity, and it is estimated as 
$\dot{M}_{ev}\sim 2\cdot 10^{16}(\zeta/0.01)(r/3\cdot 10^9{\rm cm})
(L_{sd}/5\cdot 10^{34}{\rm erg~s^{-1}}){\rm g~s^{-1}}$; that is, 
if the efficiency is $\zeta\sim 1\%$ 
(about 10\% of the gamma-ray luminosity), most of the disk matter 
from the companion star will be 
evaporated at $r\sim 3\times 10^9{\rm cm}$ and the disk will be outside 
the light cylinder, indicating the rotation powered activities of 
the pulsars are still on in the high state.  In our model,  
the X-ray emissions are produced by the intra-binary shock
 due to the interaction of the pulsar wind 
 with the outflow from the star/disk.
 
For the high state (since June 2013), we have found that the X-ray emissions 
do not show the orbital modulation (section 3.2). This would suggest that 
the shock enshrouds the pulsar and stops almost all of the pulsar wind. Since 
 the pulsar is surrounded by the disk, it would be possible that 
the outflow matter from the disk caused by  the irradiation of intense pulsar 
wind/high-energy radiation stops most of the pulsar wind.  In the present 
model, the momentum ratio of outflow matter from the disk and 
the pulsar wind may be estimated from $\eta\sim \dot{M}_{ev}v_{es}c/L_{sd}
\sim 2(\dot{M}_{ev}/10^{16}{\rm gs^{-1}})(v_{es}/3\cdot 10^8{\rm cm^{-1}})
(L_{sd}/5\cdot 10^{34}{\rm erg s^{-1}})^{-1}$, suggesting the shock 
enshrouds the pulsar. Because we do not know the exact geometry of the shock, 
which may be more complicated than a spherical shape,  we assume that 
the shock distance from the pulsar does not depend on the direction. 
In our calculation, we assumed the shock distance  
$r_s\sim 3\times10^{10}$cm and the magnetization parameter 
at the shock $\sigma\sim10^{-2}$ are the same as the shock parameters 
of the low state. To explain the increase of the 
 observed flux since June 2013,  
we assumed that $\sim 70$\% of the pulsar wind is stopped by the shock. 
We also used power index $p=2$ of the particle distribution to explain 
the hardness of the X-ray spectrum. 
 Right panel of Figure~\ref{fit} compares the 
model calculation with the multi-wavelength observations for the high state.  
The GeV gamma-ray emission in the high state is explained by
 the inverse-Compton process between the pulsar wind and UV/Optical emissions 
from the disk. The calculated total luminosity from the pulsar wind 
is $L_{pw}\sim 3\times 10^{34}{\rm erg s^{-1}}$.

The \nustar\ data of J1023 provide the first hard X-ray spectra extending to 79 keV for a transitional MSP/LMXB in both MSP and LMXB states. Together with multi-wavelength observations from radio to gamma-ray, these high-quality \nustar\ spectra allow us to constrain the high-energy emission mechanisms and we show that the data above 10 keV are crucial.  In summary,
the length of 350MHz eclipse in the low state may suggest that the
 outflow from the companion star dominates the pulsar wind and the shock 
surrounds the pulsar.  This could be  due to a large mass loss 
rate $\dot{M}>2\times 10^{16}{\rm g\, s^{-1}}$ of the stellar wind or existing 
of the  magnetic pressures of the star. In such a geometry, 
the orbital modulation of the X-rays is caused by Doppler 
boosting due to finite velocity of the shocked pulsar wind. The disappearance 
of the orbital modulation of the X-rays founded by \nustar\ during the LMXB state may indicate that  the evaporating matter from the 
disk due to the pulsar irradiation stops almost all the pulsar wind.

We note that at the time of revising this paper, a paper by the \nustar\ team using the same datasets has been recently accepted for publication (Tendulkar et al. 2014) and their results are entirely consistent with ours.

KLL, AKHK, and RJ are supported by the Ministry of Science and Technology
 of the Republic of China (Taiwan) through grants 100-2628-M-007-002-MY3, 100-2923-M-007-001-MY3, and 101-2119-M-008-007-MY3.
JT and KSC are supported by a GRF grant of HK Government
 under HKU 17300814P.  PHT is supported by the Ministry of Science and Technology of the Republic of China (Taiwan) through grant 101-2112-M-007-022-MY3. CYH is supported by the National Research Foundation of Korea through grant 2011-0023383. This research has
made use of the \nustar\ Data Analysis Software
(NuSTARDAS) jointly developed by the ASI Science
Data Center (ASDC, Italy) and Caltech (USA).

{\it Facilities:} \facility{NuSTAR}, \facility{Swift}


\begin{references}

\reference{} Alexander, D. M., Stern, D., Del Moro, A., et al. 2013, ApJ, 773,
125



\reference{} Alpar, M.A., Cheng, A.F., Ruderman, M.A.,  Shaham, J. 1982, Nature, 300, 728

\reference{} Archibald, Anne M., Kaspi, V.M., Hessels, J.W. T.,
 Stappers, B., Janssen, G.,  Lyne, A., 2013, ApJ, submitted (arXiv:1311.5161)

\reference{} Archibald, Anne M., Kaspi, Victoria M., Bogdanov, Slavko,
Hessels, Jason W. T., Stairs, Ingrid H., Ransom, Scott M.,
 McLaughlin, Maura A., 2010, 722, 88

\reference{} Archibald, A. M. et al. 2009, Science, 324, 1411

\reference{} Bassa, C.~G. et al. 2014, MNRAS, 441, 1825



\reference{} Bogdanov, S. Archibald, A.M., Hessels, J.W. T., Kaspi, V.M.,
Lorimer, D. McLaughlin, M.A., Ransom, S.M., Stairs, I.H., 2011, ApJ, 742, 97

\reference{} Bond, H.E., White, R.L., Becker, R.H., O'Brien, M.S., 2002, PASP,
 114, 1359

\reference{} Campana, S., Colpi, M., Mereghetti, S., Stella, L., Tavani, M., 1998, A\&ARv, 8, 279


\reference{} de Martino, D., Belloni, T., Falanga, M., Papitto, A., Motta, S., Pellizzoni, A., Evangelista, Y., Piano, G., Masetti, N., Bonnet-Bidaud, J.-M., Mouchet, M., Mukai, K., Possenti, A. 2013, A\&A, 550, A89



\reference{} Frank, J., King, A., \& Raine, D. 2002, Accretion Power in Astrophysics (Cambridge: Cambridge Univ. Press)

\reference{} Ferrigno, C., Bozzo, E., Papitto, A., Rea, N., Pavan, L., Campana, S., Wieringa, M., Filipovic, M., Falanga, M., Stella, L. 2013, A\&A, in press, arXiv:1310.7784


 Bogdanov, S., 2013, ATel, 5514, 1

\reference{} Harrison, F. A., Craig, W. W., Christensen, F. E., et al. 2013,
ApJ, 770, 103

\reference{} Homer, L., Szkody, P., Chen, B., Henden, A., Schmidt, G., Anderson,\
 S.F., Silvestri, N.M., Brinkmann, J., 2006, AJ, 131, 562
 
 \reference{} Hui, C.~Y., Tam, P.~H.~T., Takata, J., Kong, A.~K.~H., Cheng, K.~S., Wu, J.~ H.~K., Lin, L.~C.~C., Wu, E.~M.~H. 2014, ApJ, 781, L21


\reference{} Kong, A.K.H., 2013, ATel, 5515, 1

\reference{} Kong, S.W., Cheng, K.S., Huang, Y.F., 2012, ApJ, 753, 127



\reference{} Linares, M., Bahramian, A., Heinke, C., Wijnands, R., Patruno, A., Altamirano, D., Homan, J., Bogdanov, S., Pooley, D. 2013, MNRAS, 438, 251




\reference{} Papitto, A., Ferrigno, C., Bozzo, E., et al. 2013, Nature, 501, 517

\reference{} Papitto, A. Torres, D., Li, J. 2014, MNRAS, 438, 2106

\reference{} Patruno, A., Archibald, A. M., Hessels, J. W. T., Bogdanov, S., Stappers, B. W., Bassa, C. G., Janssen, G. H., Kaspi, V. M., Tendulkar, S., Lyne, A. G. 2014, ApJ, 781, L3

\reference{} Romanova, M. M., Ustyugova, G. V., Koldoba, A. V., \& Lovelace, R. V. E. 2009, MNRAS, 399, 1802


\reference{} Stappers, B.W. et al. 2014, ApJ, submitted, arXiv:1311.7506

\reference{} Tam, P.H.T., Hui, C.Y., Huang, R.H.H., Kong, A.K.H., Takata, J.,
Lin, L.C.C., Yang, Y.J., Cheng, K.S., Taam, R. E., 2010, ApJ, 724, L207

\reference{} Takata, J., Cheng, K.S., Taam, R.E., 2012, ApJ, 745, 100

\reference{} Takata, J., Cheng, K.S., Taam, R.E., 2010, ApJ, 723, L68

\reference{} Takata, J., Li, K. L., Leung, G. C. K., et al. 2014, ApJ, 785, 131

\reference{} Tendulkar, S.P. et al. 2014, ApJ, 791, 77

\reference{} Thorstensen, J.R., Armstrong, E., 2005, AJ, 130, 759


\reference{} Wang, Z., Archibald, A.M., Thorstensen, J.R., Kaspi, V.M.,
Lorimer, D.R., Stairs, I., Ransom, S.M., 2009, ApJ, 703, 2017




\end{references}
\end{document}